\documentclass[twocolumn,showpacs,secnumarabic,superscriptaddress,amsmath,amssymb,aps,pra]{revtex4-1}
\usepackage{graphicx}
\usepackage{dcolumn}
\usepackage{amsmath,bm}
\usepackage{epstopdf}
\usepackage[mathlines]{lineno}
\usepackage{color}
\usepackage[colorlinks=true,linkcolor=blue,citecolor=magenta,urlcolor=cyan]{hyperref}

\begin{document}
\title{Hyperfine interaction mediated electric-dipole spin resonance: the role of frequency modulation}
\author{Rui\! Li}
\email{rl.rueili@gmail.com}
\affiliation{Quantum Physics and Quantum Information Division, Beijing Computational Science Research Center, Beijing 100193, People's Republic of China}

\begin{abstract}
The electron spin in a semiconductor quantum dot can be coherently controlled by an external electric field, an effect called electric-dipole spin resonance (EDSR). Several mechanisms can give rise to the EDSR effect, among which there is a hyperfine mechanism, where the spin-electric coupling is mediated by the electron-nucleus hyperfine interaction. Here, we investigate the influence of frequency modulation (FM) on the spin-flip efficiency. Our results reveal that FM plays an important role in the hyperfine mechanism. Without FM, the electric field almost cannot flip the electron spin; the spin-flip probability is only about 20\%. While under FM, the spin-flip probability can be improved to approximately 70\%. In particular, we find that the modulation amplitude has a lower bound, which is related to the width of the fluctuated hyperfine field.
\end{abstract}
\date{\today}
\maketitle

\section{Introduction}
The electron spin confined in a semiconductor quantum dot is a promising candidate for the qubit~\cite{Hanson,Buluta}. Since the pioneering work of Loss and DiVincenzo in 1998~\cite{Loss}, spin quantum computing has become an important research area. There have been many theoretical and experimental advances in the subsequent decades. Qubit initialization~\cite{Johnson,Petta}, coherent manipulation~\cite{Koppens,Foletti}, and single shot readout~\cite{Elzerman,Barthel,Morello}  can be achieved with high precision. The quantum dot spin qubit has many advantages. For example, spin has a relative long coherence time (in comparison with the qubit manipulation time)~\cite{Bluhm}, the two-qubit manipulation can be easily achieved by using the Heisenberg exchange interaction~\cite{Burkard,Xuedong,RuiLi5,RuiLi4}, and it is more or less much easier to achieve the qubit scalability.

Single spin manipulation can be achieved by using the traditional electron spin resonance technique~\cite{Koppens}. This method not only needs a static Zeeman field, but also needs a perpendicular ac magnetic field~\cite{Awschalom,Meisels}. When the frequency of the ac field matches the electron Zeeman splitting, one achieves coherent spin rotation. However, it is difficult to produce a local ac magnetic field experimentally~\cite{Koppens}. Recently, instead of an ac magnetic field, an ac electric field has been used to manipulate a single electron spin, an effect called electric-dipole spin resonance (EDSR)~\cite{Tokura,Pioro,Nowack,Nadj1,Nadj2,Rashba,Golovach,RuiLi,RuiLi3,Laird,Rashba3}. It is much easier to produce a local ac electric field in experiments.

We can explain quantum dot EDSR in the following simple way. In absence of electric driving, the electron's spin degree of freedom is mixed with its orbital degree of freedom due to some mechanisms, e.g., the slanted magnetic field~\cite{Yoneda,Palyi,Szechenyi,Scarlino}, spin-orbit coupling~\cite{Ban,Yang,Nowak,Khomitsky,Echeverria,Romhanyi,ZhangPeng}, and the electron-nucleus hyperfine interaction~\cite{Shafiei,Osika,Chesi}. Therefore, an ac electric field can induce electric-dipole transitions between the electron Zeeman levels.

The hyperfine interaction mediated EDSR was first observed experimentally in 2007~\cite{Laird}. The hallmarks of this mechanism are listed as follows. First, under electric driving, there is no Rabi oscillation, i.e., no periodical revival of the spin polarization~\cite{Laird}; the spin-flip probability monotonically increases to a saturating value. Second, the Rabi frequency is independent of the Zeeman field~\cite{Laird}. Third, if there is no frequency modulation (FM) of the driving electric field, the electron spin cannot be flipped~\cite{Shafiei}. The first theoretical investigation into the hyperfine mechanism was given by Rashba~\cite{Rashba3}, who provides deep insight into the physics behind the EDSR effect. However, the developed theory was built on the mean field approximation, and, in particular, the effects of FM were not explored.

In this paper, we re-address the mechanism of hyperfine interaction mediated EDSR in the presence of FM. Our main findings are summarized as follows. First, we derive an analytical expression for the spin-flip probability where we do not take any mean field approximation. The spin-flip probability is expressed by a summation over a large number of periodic functions such that the oscillating behaviour of each periodic function is covered by the summation. Second, under fixed-frequency driving, the spin cannot be flipped. Because the electron-nucleus hyperfine interaction brings an inhomogeneous broadening to the spin splitting; the fixed driving frequency cannot match the inhomogeneously broadened spin splitting. Third, FM is used to broaden the frequency spectrum of the driving electric field. When the width of the frequency spectrum is larger than the width of the fluctuated hyperfine field, the spin-flip probability can be greatly improved.


\section{\label{Sec_II}EDSR in a quantum dot with quasi-1D confinement}
In order to explicitly show the underlying physics of hyperfine interaction mediated EDSR, we first study this phenomena in a simple quasi-1D quantum dot. In fact, a quantum dot with quasi-1D  confinement, e.g., an InAs~\cite{Schroer,Nadj1} or an InSb~\cite{Nadj2} nanowire quantum dot can already be fabricated experimentally. Therefore, we confront the following total Hamiltonian
\begin{eqnarray}
H_{\rm tot}&=&\frac{p^{2}_{x}}{2m_{e}}+\frac{1}{2}m_{e}\omega^{2}_{0}x^{2}+\sum^{N}_{l=1}\frac{A}{\hbar^{2}}{\bf S}\cdot{\bf I}_{l}\delta(x-x_{l})\nonumber\\
&&+\gamma_{e}BS_{x}+eEx\cos\left[\int^{t}_{0}dt'\nu(t')\right],\label{EQ_originalHamiltonian}
\end{eqnarray}
where $p_{x}=-i\hbar\partial_{x}$, $m_{e}$ is the effective electron mass, $\hbar\omega_{0}$ is the orbital energy of the quantum dot, $A$ is the strength of the electron-nucleus hyperfine coupling, $\gamma_{e}$ is the electron gyromagnetic ratio, ${\bf S}$ ($S=1/2$) and ${\bf I}_{l}$ ($I=1/2$) are the electron and the $l$-th nuclear spin operators, respectively, $N$ is the total number of the nuclear spins in the dot, $x_{l}$ is the site of the $l$-th nuclear spin, and $\nu\,(t')$ is the modulated frequency of the driving electric field.

We have introduced the concept of FM in Eq.~(\ref{EQ_originalHamiltonian}), where the frequency of the driving field is time dependent. In this paper, we will study both the fixed-frequency driving case, i.e., $\nu(t')=\nu_{0}$, and the modulated-frequency driving case, i.e.,  $\nu(t')=\nu_{0}+\delta\nu\cos(\nu_{\rm fm}t')$, where $\delta\nu$ and $\nu_{\rm fm}$ are the modulation amplitude and the modulation frequency, respectively.

\subsection{The mixing of spin and orbital degrees of freedom owing to hyperfine interaction}
Our first step is to calculate the spin splitting and the corresponding wavefunctions in the quantum dot in the presence of both an external magnetic field and a hyperfine field. The Hamiltonian of the quantum dot in the absence of the driving term can be divided into two parts
\begin{eqnarray}
H&=&H_{0}+H_{1},\nonumber\\
H_{0}&=&\frac{p^{2}_{x}}{2m_{e}}+\frac{1}{2}m_{e}\omega^{2}_{0}x^{2}+\gamma_{e}BS_{x},\nonumber\\
H_{1}&=&\sum^{N}_{l=1}\frac{A}{\hbar^{2}}\left[S_{x}I^{x}_{l}+\frac{1}{2}(S_{+}I^{-}_{l}+S_{-}I^{+}_{l})\right]\delta(x-x_{l}),
\end{eqnarray}
where $S_{\pm}=S_{y}\pm\,iS_{z}$ and $I^{\pm}_{l}=I_{y}\pm\,iI_{z}$ are the electron and the $l$-th nuclear spin raising/lowering operators, respectively. Note that the spin quantization direction here is along the $x$-axis. The full quantum mechanical solution to the time-independent Schr\"odinger equation is impossible because the hyperfine term $H_{1}$ is coordinate-dependent.  However, when the Zeeman field $\gamma_{e}B$ is strong enough, i.e., the Zeeman field is much larger than the hyperfine field, we can solve the energy spectrum by treating $H_{1}$ as a perturbation. The related physical quantities should satisfy the following perturbation condition:
\begin{equation}
\sum^{N}_{l=1}(A_{l}/4)\ll\gamma_{e}B\ll\hbar\omega_{0},
\end{equation}
where $A_{l}$ is the $l$-th electron-nuclear hyperfine coupling coefficient. The definition of this coefficient is given in a later equation~(\ref{Eq_hyperfine_coeff}).

It is easy to find the zeroth order eigenvalues and the corresponding eigenfunctions:
\begin{eqnarray}
E^{0}_{n\sigma_{e}\chi_{m}}&=&(n+1/2)\hbar\omega_{0}+(1/2)(-1)^{\sigma_{e}}\gamma_{e}B,\nonumber\\
|\Psi^{0}_{n\sigma_{e}\chi_{m}}\rangle&=&\psi_{n}(x)|\sigma_{e}\rangle\otimes|\chi_{m}\rangle,\label{Eq_energy_zero}
\end{eqnarray}
where $n=0,1,2\ldots$ is the main quantum number, $\sigma_{e}=0,1$ is the electron spin quantum number, with $|0\rangle\rightarrow|\!\uparrow_{x}\rangle$ and $|1\rangle\rightarrow|\!\downarrow_{x}\rangle$, $|\chi_{m}\rangle=|\sigma^{m}_{1}\sigma^{m}_{2}\cdots\sigma^{m}_{N}\rangle$ is the $N$-bit binary representation of a decimal number $m=0,1,\ldots(2^{N}-1)$, with $\sigma^{m}_{l}=0,1$ being the $l$-th nuclear spin quantum number, and $\psi_{n}(x)$ is the eigenfunction of the harmonic oscillator, e.g.,
\begin{eqnarray}
\psi_{0}(x)&=&1/(\pi^{1/4}x^{1/2}_{0}){\rm exp}\left[-x^{2}/(2x^{2}_{0})\right],\nonumber\\
\psi_{1}(x)&=&2^{1/2}/(\pi^{1/4}x^{1/2}_{0})(x/x_{0}){\rm exp}\left[-x^{2}/(2x^{2}_{0})\right],
\end{eqnarray}
with $x_{0}=\sqrt{\hbar/(m_{e}\omega_{0})}$ being the characteristic length of the quantum dot. The electron in the quantum dot should occupy the lowest orbital level, i.e., $n=0$. Therefore, in the following we only focus on $n=0$ Zeeman levels $|\Psi_{0\sigma_{e}\chi_{m}}\rangle$.

\begin{figure}
\centering
\includegraphics{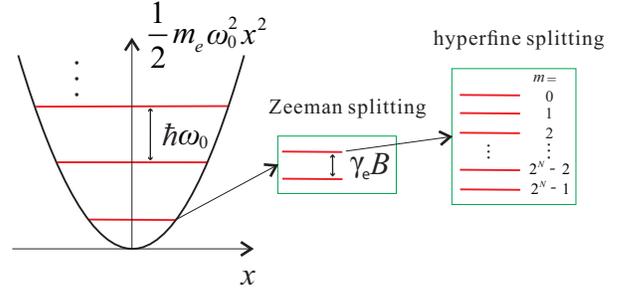}
\caption{\label{Fig_energyspectrum}The energy spectrum of the quantum dot in the presence of the electron-nucleus hyperfine interaction.}
\end{figure}
It should be noted that, up to the zeroth order, each energy level has a degeneracy of $2^{N}$ (see equation~(\ref{Eq_energy_zero})). However, the perturbation $H_{1}$ does not mix the states in the degenerate subspace. Therefore, in our following calculation we can use the non-degenerate perturbation formulas~\cite{Landau}
\begin{eqnarray}
E&=&E^{0}+\langle\Psi^{0}|H_{1}|\Psi^{0}\rangle,\nonumber\\
|\Psi\rangle&=&|\Psi^{0}\rangle+\sum_{\Psi'}\frac{\langle\Psi'^{0}|H_{1}|\Psi^{0}\rangle}{E^{0}-E'^{0}}|\Psi'^{0}\rangle.
\end{eqnarray}
Up to the first-order perturbation, the Zeeman energies have the following rectifications:
\begin{eqnarray}
E_{00\chi_{m}}&=&(1/2)\hbar\omega_{0}+(1/2)\gamma_{e}B+\sum^{N}_{l=1}(A_{l}/4)(-1)^{\sigma^{m}_{l}},\nonumber\\
E_{01\chi_{m}}&=&(1/2)\hbar\omega_{0}-(1/2)\gamma_{e}B-\sum^{N}_{l=1}(A_{l}/4)(-1)^{\sigma^{m}_{l}},\label{Eq_hyperfine_coeff}
\end{eqnarray}
where $A_{l}=A\psi^{2}_{0}(x_{l})$ is the $l$-th electron-nucleus hyperfine coupling coefficient. The hyperfine energy structure of the quantum dot is schematically shown in figure~\ref{Fig_energyspectrum}. The corresponding first-order perturbation wavefunctions read
\begin{eqnarray}
|\Psi_{00\chi_{m}}\rangle&=&|\Psi^{0}_{00\chi_{m}}\rangle+\sum^{N}_{l=1}\frac{A\psi_{1}(x_{l})\psi_{0}(x_{l})}{2(\gamma_{e}B-\hbar\omega_{0})}\sigma^{m}_{l}\nonumber\\
&&\times\psi_{1}(x)|1\rangle\otimes
|\sigma^{m}_{1}\ldots\sigma^{m}_{l-1}0\sigma^{m}_{l+1}\ldots\sigma^{m}_{N}\rangle,\nonumber\\
|\Psi_{01\chi_{m}}\rangle&=&|\Psi^{0}_{01\chi_{m}}\rangle+\sum^{N}_{l=1}\frac{A\psi_{1}(x_{l})\psi_{0}(x_{l})}{2(-\gamma_{e}B-\hbar\omega_{0})}\frac{(-1)^{\sigma^{m}_{l}}+1}{2}\nonumber\\
&&\times\psi_{1}(x)|0\rangle\otimes|\sigma^{m}_{1}\ldots\sigma^{m}_{l-1}1\sigma^{m}_{l+1}\ldots\sigma^{m}_{N}\rangle.\label{Eq_wavefunction}
\end{eqnarray}
Thus, it can be seen clearly from equation~(\ref{Eq_wavefunction}) that, owing to the hyperfine interaction, the spin and the orbital degrees of freedom of the electron are mixed. Therefore, it is expected that the electric field will induce the transitions between the up Zeeman level and the down Zeeman level.

\subsection{The spin-flip probability under fixed-frequency driving}
In this subsection, we study the EDSR effect under fixed-frequency driving. The driving Hamiltonian is written as $eEx\cos(\nu_{0}t)$, where $\nu_{0}$ is the driving frequency. Although the electron spin splitting is broadened by the hyperfine interaction (see equation~(\ref{Eq_hyperfine_coeff}) and figure~\ref{Fig_energyspectrum}), we still choose the driving frequency to satisfy the `resonant' condition, i.e., $\hbar\nu_{0}=\gamma_{e}B$.

As we have shown in figure~\ref{Fig_energyspectrum}, each Zeeman (both the up and the down) level has been broadened to a series of sublevels, where $m$ is the sublevel index. So what is the initial state for the coupled electron-nuclear system? Because the nuclear Zeeman splitting is much less than the Boltzmann energy $k_{B}T$, with $T\sim\,m$K being the temperature of the experiments. It is a good approximation to assume that the nuclear spin ensemble is totally unpolarized.  In other words,  in the initial mixed state of the total electron-nucleus system, each pure state $|\Psi_{00\chi_{m}}\rangle$ has the equal probability
\begin{equation}
\rho(0)=(1/2^{N})\sum^{2^{N}-1}_{m=0}|\Psi_{00\chi_{m}}\rangle\langle\Psi_{00\chi_{m}}|.\label{Eq_initialstate}
\end{equation}
Here, we have let the electron initially be in the spin-up state $|0\rangle$ ($|\uparrow_{x}\rangle$).

\begin{figure}
\centering
\includegraphics{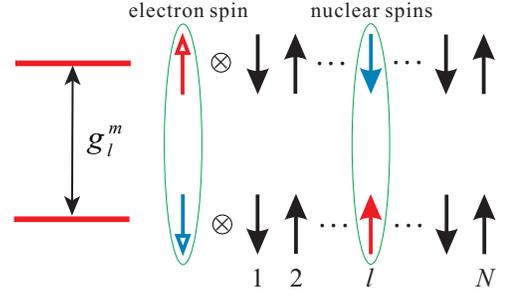}
\caption{\label{Fig_flipflop}Schematic diagram of the electric-dipole transition between the state $|\Psi_{00\chi_{m}}\rangle$  and the state $|\Psi_{01\chi'_{m}}\rangle$. In other words, the driving electric field assists the electron-nucleus spin-flip-flop process. There is an exchange of spin polarization between the electron and the $l$-th nucleus.}
\end{figure}

Under electric driving, we want to calculate the electron spin-flip rate for the initial mixed state given in equation~(\ref{Eq_initialstate}). Our first step is to calculate the spin-flip rate for the initial pure state $|\Psi_{00\chi_{m}}\rangle$.  From equation~(\ref{Eq_wavefunction}), we can calculate the electric-dipole transition element between the state $|\Psi_{00\chi_{m}}\rangle$ and the state $|\Psi_{01\chi'_{m}}\rangle$
\begin{eqnarray}
g^{m}_{l}&=&\langle\Psi_{00\chi_{m}}|eEx|\Psi_{01\chi'_{m}}\rangle\nonumber\\
&=&-\frac{eEx_{0}\times\hbar\omega_{0}}{\sqrt{2}\left(\hbar^{2}\omega^{2}_{0}-\gamma^{2}_{e}B^{2}\right)}\sigma^{m}_{l}A\psi_{1}(x_{l})\psi_{0}(x_{l})\nonumber\\
&\approx&-\frac{eEx_{0}}{\sqrt{2}\hbar\omega_{0}}\sigma^{m}_{l}A\psi_{1}(x_{l})\psi_{0}(x_{l}),\label{Eq_transitionrate}
\end{eqnarray}
where $|\chi_{m}\rangle=|\sigma^{m}_{1}\cdots\sigma^{m}_{l}\cdots\sigma^{m}_{N}\rangle$ and $|\chi'_{m}\rangle\equiv|\sigma^{m}_{1}\cdots\sigma^{m}_{l-1}0\sigma^{m}_{l+1}\cdots\sigma^{m}_{N}\rangle$. Similar results were also obtained in~\cite{Rashba3,Rudner}. The above equation tells us that the electron-nucleus spin-flip-flop process is assisted by the driving electric field (see figure~\ref{Fig_flipflop}). There is an exchange of the spin polarization between the electron and the $l$-th nucleus. It should be noted that the value of $\sigma^{m}_{l}$ in equation~(\ref{Eq_transitionrate}) automatically guarantees the feasibility of this flip-flop process, i.e., only when $\sigma^{m}_{l}=1$ is $g^{m}_{l}\neq0$. The energy difference between these two states reads
\begin{equation}
E_{00\chi_{m}}-E_{01\chi'_{m}}\approx\gamma_{e}B+\sum^{N}_{l=1}(A_{l}/2)(-1)^{\sigma^{m}_{l}}.\label{Eq_broadenZeeman}
\end{equation}
There arises another problem: under electric driving there are many quantum states with which the state $|\Psi_{00\chi_{m}}\rangle$ can flip-flop, as long as the $l$-th bit in $|\chi_{m}\rangle$ has the value $\sigma^{m}_{l}=1$ $(l\in\{1,\ldots\,N\})$. Despite this, the spin-flip probability for the initial pure sate $|\Psi_{00\chi_{m}}\rangle$ still has a simple expression (for details see appendix~\ref{Appendix_A})
\begin{equation}
P^{m}_{\downarrow_{x}}(t)=\frac{\sum^{N}_{l=1}(g^{m}_{l})^{2}}{\hbar^{2}\Omega^{2}_{m}}\sin^{2}\frac{\Omega_{m}t}{2},\label{Eq_inversion_purestate}
\end{equation}
where
\begin{subequations}
\begin{eqnarray}
\Delta_{m}&=&-\sum^{N}_{l=1}(A_{l}/2)(-1)^{\sigma^{m}_{l}},\label{Eq_detuning}\\
\Omega_{m}&=&\frac{1}{\hbar}\left[\Delta^{2}_{m}+\sum^{N}_{l=1}(g^{m}_{l})^{2}\right]^{1/2}.\label{Eq_inversion_parameters}
\end{eqnarray}
\end{subequations}
Here, $\Delta_{m}$ is the detuning introduced by the hyperfine field and $\Omega_{m}$ is the Rabi frequency in the presence of the detuning. Therefore, for the initial pure state $|\Psi_{00\chi_{m}}\rangle$, we obtain a simple Rabi formula~\cite{Scully} for the spin-flip probability. As we can see from equations~(\ref{Eq_transitionrate}) and (\ref{Eq_inversion_parameters}), the Rabi frequency $\Omega_{m}$ is independent of the magnetic field, which is consistent with experimental observations~\cite{Laird,Shafiei}.

Since we have already obtained the spin-flip probability for the initial pure state, the generalization to the initial mixed state~(\ref{Eq_initialstate}) is simple. The spin-flip probability for the initial mixed state reads as
\begin{equation}
P_{\downarrow_{x}}(t)=(1/2^{N})\sum^{2^{N}-1}_{m=0}\frac{\sum^{N}_{l=1}(g^{m}_{l})^{2}}{\hbar^{2}\Omega^{2}_{m}}\sin^{2}\frac{\Omega_{m}t}{2}.\label{Eq_inversion}
\end{equation}
This is just a summation over all the pure state's spin-flip probability. Because equation~(\ref{Eq_inversion}) is summed over a large number of periodic functions, $P_{\downarrow_{x}}(t)$ may not contain the oscillating behaviour when $N$ is very large. 

\begin{figure}
\centering
\includegraphics{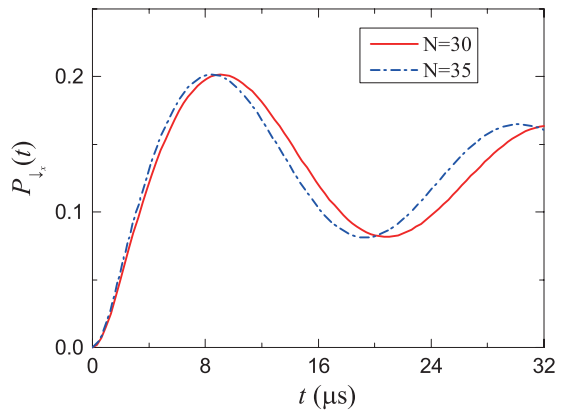}
\caption{\label{Fig_EDSRWithoutFM}The spin down probability under the fixed-frequency driving. The quantum dot parameters are chosen as $x_{0}=100$ nm, $A/(\hbar\pi^{1/2}x_{0})=0.5$ MHz, and $eEx_{0}/(\hbar\omega_{0})=0.2$.  $N$ is the total number of the nuclear spins in the quantum dot and the nuclear spins are equal-distance distributed in the quantum dot.}
\end{figure}

In figure~(\ref{Fig_EDSRWithoutFM}), we show the spin-flip probability as a function of the driving time. As we can see, under fixed frequency driving ($\hbar\nu_{0}=\gamma_{e}B$), the electron spin almost can not be flipped;  the maximal spin-down probability is only $0.2$. This very low spin-flip probability has not been observed in experiment~\cite{Shafiei}. Therefore, under fixed-frequency driving, EDSR actually can not occur. We will explain why the spin-flip probability is very low in the next subsection.

\subsection{The spin-flip probability under modulated-frequency driving}
As we have explicitly shown in the above subsection, each pure state $|\Psi_{00\chi_{m}}\rangle$ in the initial mixed state (see equation~(\ref{Eq_initialstate})) brings a detuning $\Delta_{m}$ to the spin-flip probability $P_{\downarrow_{x}}(t)$ (see equations~(\ref{Eq_detuning}) and (\ref{Eq_inversion})). When $m$ runs one by one from $0$ to $2^{N}-1$, the value of the detuning $\Delta_{m}$ has a distribution function. Actually, this distribution describes the fluctuation of the hyperfine field. The distribution function approximately reads~\cite{Merkulov}
\begin{equation}
P(\Delta)=\frac{1}{(2\pi)^{1/2}\Delta_{\rm flu}}e^{-\Delta^{2}/(2\Delta^{2}_{\rm flu})},\label{Eq_distribution}
\end{equation}
where $\Delta_{\rm flu}=(1/2)\sqrt{\sum^{N}_{l=1}A^{2}_{l}}$ is the width of the hyperfine field fluctuation. Therefore, under fixed-frequency driving (e.g., $\gamma_{e}B=\hbar\nu_{0}$), the detuning $\Delta$ is not fixed. This is the essential reason that leads to the very low spin-flip probability. 

\begin{figure}
\includegraphics{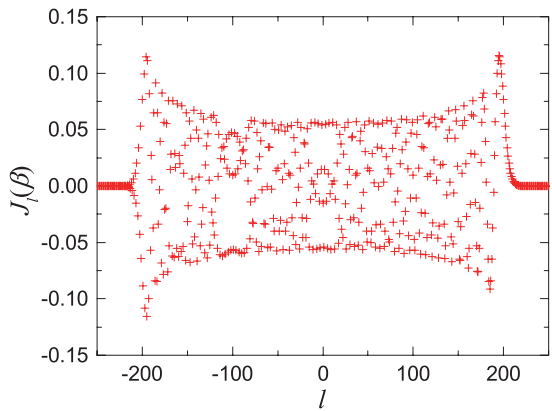}
\caption{\label{Fig_Bessel}The Bessel function $J_{l}(\beta)$ as a function of the index $l$ for large modulation index $\beta=200$. For $|l|>\beta$, $J_{l}(\beta)\approx0$.}
\end{figure}

FM is a concept in telecommunication and signal processing. As we will show in the following, FM can solve the above very low spin-flip probability problem. Here, we take the simplest example to explain the importance of the FM. We consider the instantaneous frequency of the driving electric field is modulated to the following form
\begin{equation}
\nu(t')=\nu_{0}+\delta\nu\cos(\nu_{\rm fm}t'),
\end{equation}
where $\nu_{0}$ is the central frequency, $\delta\nu$ is the modulation amplitude, and $\nu_{\rm fm}$ is the modulation frequency. In order to understand FM straightforwardly, we also give the experimental parameters here, where $\nu_{0}=2.9$ GHz, $\delta\nu=36$ MHz, and $\nu_{\rm fm}=3$ kHz~\cite{Laird}. In the presence of FM, the driving Hamiltonian can be written as~\cite{Haykin}
\begin{equation}
eEx\cos\left[\int^{t}_{0}dt'\nu(t')\right]=\sum^{\infty}_{l=-\infty}eExJ_{l}(\beta)\cos\left[(\nu_{0}+l\nu_{\rm fm})t\right],\label{Eq_FMdriving}
\end{equation}
where we have made a Fourier transformation, $\beta=\delta\nu/\nu_{\rm fm}\gg1$ is the modulation index (we only consider the wideband modulation here), 
 and
\begin{equation}
J_{l}(\beta)=\frac{1}{2\pi}\int^{\pi}_{-\pi}dx\,{\rm exp}[i(\beta\sin\,x-lx)]\nonumber
\end{equation}
is the $n$-th order Bessel function of the first kind. In figure~{\ref{Fig_Bessel}}, we show the Bessel function as a function of the index $l$ for the large modulation index $\beta=200$. As we can see, when $|l|>\beta$, $J_{l}(\beta)\approx0$. Therefore, actually, the summation of $l$ in equation~(\ref{Eq_FMdriving}) is constrained to $-\beta\leq\,l\leq\beta$.

\begin{figure}
\centering
\includegraphics{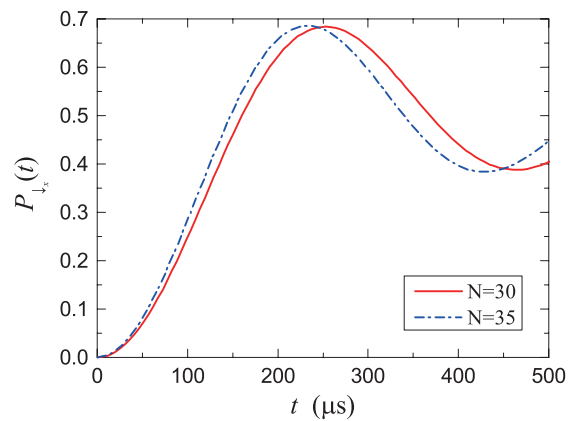}
\caption{\label{Fig_EDSRFM}The spin down probability under modulated frequency driving. The modulation parameters are chosen as $\beta=200$ and $\nu_{\rm fm}=0.02$ MHz. The other parameters are the same as in figure~\ref{Fig_EDSRWithoutFM}.}
\end{figure}

From equation~(\ref{Eq_FMdriving}), it is easy to find that the frequency spectrum of the driving electric field has been broadened by the FM. The frequency of the driving now can run from $\nu_{0}-\delta\nu$ to $\nu_{0}+\delta\nu$, where $\delta\nu=\beta\nu_{\rm fm}$ can be considered as the width of the frequency spectrum. This frequency spectrum broadening is very useful. As we have shown in equation~(\ref{Eq_distribution}), the detuning $\Delta$, i.e., the hyperfine field, also has a distribution which approximately ranges from $-\sqrt{2}\Delta_{\rm flu}$ to $\sqrt{2}\Delta_{\rm flu}$. Therefore, the inhomogeneously broadened spin splitting is ranged from $\gamma_{e}B-\sqrt{2}\Delta_{\rm flu}$ to $\gamma_{e}B+\sqrt{2}\Delta_{\rm flu}$. When the width of the frequency spectrum $\delta\nu$ is larger than the width of the fluctuating hyperfine field $\sqrt{2}\Delta_{\rm flu}$, i.e.,
\begin{equation}
\sqrt{2}\Delta_{\rm flu}<\delta\nu,
\end{equation}
the detuing $\Delta$ will not appear in the spin-flip probability $P_{\downarrow_{x}}(t)$. Because for each initial pure state $|\Psi_{00\chi_{m}}\rangle$ the broadened spin splitting is $\gamma_{e}B-\Delta_{m}$ (see equation~(\ref{Eq_broadenZeeman})), we can always find a special mode $l$ in the driving field that matches this splitting, where
\begin{equation}
\gamma_{e}B-\Delta_{m}=\nu_{0}+l\nu_{\rm fm}.
\end{equation}
Thus, the $l$-th mode $eExJ_{l}(\beta)\cos\left[(\nu_{0}+l\nu_{\rm m})t\right]$, where $l=-\Delta_{m}/\nu_{\rm fm}$, is in resonance with this broadened spin splitting $\gamma_{e}B-\Delta_{m}$. Therefore, the spin-flip probability under the FM can be written as
\begin{equation}
P_{\downarrow_{x}}(t)=\frac{1}{2^{N}}\sum^{2^{N}-1}_{m=0}\sin^{2}\frac{\Omega^{\rm fm}_{m}t}{2},\label{Eq_inversion_FM}
\end{equation}
where $\Omega^{\rm fm}_{m}=\frac{1}{\hbar}J_{-(\Delta_{m}/\nu_{\rm fm})}(\beta)\sqrt{\sum^{N}_{l=1}(g^{m}_{l})^{2}}$ is the Rabi frequency under the FM. When the detuning $\Delta_{m}$ is absent from the Rabi frequency, the spin-flip probability can been greatly improved. In figure~{\ref{Fig_EDSRFM}}, we show the spin-flip probability as a function of the driving time under FM. We find that the maximal spin-flip probability increases from 0.2 (see figure~\ref{Fig_EDSRWithoutFM}) to 0.7 (see figure~\ref{Fig_EDSRFM}). This result is consistent with the experimental observation that the EDSR can only be observed under FM~\cite{Shafiei}. We also mention that the experimentally observed maximal spin-flip probability is also about 0.7~\cite{Shafiei}.

Because the spin-flip probability $P_{\downarrow_{x}}(t)$ is still expressed by a summation over a large number of periodical functions (see equation~(\ref{Eq_inversion_FM})), the Rabi oscillation may not appear when $N$ is larger.  If $N$ is not  large enough, e.g., $N=30$ or $35$, as we show in figures~\ref{Fig_EDSRWithoutFM} and \ref{Fig_EDSRFM}, there is still an oscillation for the spin-flip probability, but the amplitude of the oscillation is small.

\section{\label{Sec_III}EDSR in quantum dot with quasi-2D confinement}
Most gated quantum dots fabricated from semiconductor heterostructures, e.g., GaAs/AlGaAs heterostructures, are confined in 2D. Here we move to consider the EDSR effect in a quantum dot with quasi-2D confinement. As we have already explored the physics of EDSR in a quasi-1D quantum dot, the generalization from 1D to 2D is natural. As we show in the following, the underlying physics does not change, except the calculations are a little more complicated. We focus on the following Hamiltonian
\begin{eqnarray}
H_{\rm tot}&=&\frac{p^{2}_{x}+p^{2}_{y}}{2m_{e}}+\frac{m_{e}\omega^{2}_{0}(x^{2}+y^{2})}{2}+\sum_{l}\frac{A}{\hbar^{2}}\textbf{S}\cdot\textbf{I}_{l}\delta(\textbf{r}-\textbf{r}_{l})\nonumber\\
&&+\gamma_{e}BS_{x}+eEx\cos\left[\int\,dt'\nu(t')\right],\label{Eq_Hamiltonian2D}
\end{eqnarray}
where $p_{x,y}=-i\hbar\partial_{x,y}$, $\textbf{r}=x\hat{e}_{x}+y\hat{e}_{y}$, and $\textbf{r}_{l}$ is the site of the $l$-th nuclear spin. The in-plane magnetic field is applied along the $x$ direction, the ac electric-field is also applied along this direction. This is the simplest Hamiltonian that captures the main physics of EDSR in a quantum dot with quasi-2D confinement.

The calculations here are completely similar to the 1D case; the dimension $y$ in equation~(\ref{Eq_Hamiltonian2D}) only introduces the calculation complexities. The transition element defined in equation~(\ref{Eq_transitionrate}) for the 2D case is modified to 
\begin{equation}
g^{m}_{l}\approx-\frac{eEx_{0}}{\sqrt{2}\hbar\omega}\sigma^{m}_{l}A\psi^{2}_{0}(y_{l})\psi_{1}(x_{l})\psi_{0}(x_{l}).\label{Eq_2Dtransitionelement}
\end{equation}
Also, the hyperfine coupling coefficient for the quantum dot with 2D confinement is defined as 
\begin{equation}
A_{l}=A\psi^{2}_{0}(x_{l})\psi^{2}_{0}(y_{l}).\label{Eq_2Dhyperfinecoefficient}
\end{equation}
Under fixed-frequency driving, the expression of the spin-flip probability is exactly the same as that given by equations~(\ref{Eq_inversion_purestate})-(\ref{Eq_inversion}). Also, under FM, the expression of the spin-flip probability still has the form shown in  equation~(\ref{Eq_inversion_FM}). For a realistic gated GaAs quantum dot, the total number of the nuclear spins in the quantum dot is about $N=10^{5}$. This number is too huge. Therefore, it is impossible to calculate the spin-flip probability $P_{\downarrow_{x}}(t)$ via equations~(\ref{Eq_inversion}) and (\ref{Eq_inversion_FM}). In the following, we only discuss some general properties of EDSR in a quasi-2D quantum dot. These properties are closely related to the experimental observations. 

Firstly, for a quantum dot with quasi-2D confinement, the Rabi frequency is still independent of the external magnetic field. As we can see from equations (\ref{Eq_inversion_parameters}) and (\ref{Eq_inversion_FM}), where $g^{m}_{l}$ and $A_{l}$ are defined in equations~(\ref{Eq_2Dtransitionelement}) and (\ref{Eq_2Dhyperfinecoefficient}), respectively, both $\Omega_{m}$  and $\Omega^{\rm fm}_{m}$ do not depend on the magnetic field strength $B$. This is consistent with experimental observations~\cite{Laird,Shafiei}.

Secondly, for a gated GaAs quantum dot, the width of the hyperfine field fluctuation is about $\Delta_{\rm flu}\approx24$ MHz (0.1 $\mu$eV)~\cite{Merkulov}. This quantity sets a lower bound on the modulation amplitude:
\begin{equation}
\sqrt{2}\Delta_{\rm flu}\approx34~{\rm MHz}<\delta\nu.
\end{equation}
In order to observe the EDSR more efficiently, it is better to let the modulation amplitude be larger than $34$ MHz.  This result is also consistent with experimental observations. An early experiment used a 36 MHz modulation amplitude~\cite{Laird} and a more recent experiment used both 40 and 75 MHz modulation amplitudes~\cite{Shafiei}.

\section{\label{Sec_IV}Summary}
The electron-nucleus hyperfine interaction, which commonly exists in III-V semiconductor quantum dots, is often considered as a nuisance because it can lead to spin decoherence~\cite{Khaetskii,Coish,Witzel,Yao,Deng,Cywinski,RuiLi2,Echeverria2}. On the other hand, we can make use of the hyperfine interaction. The hyperfine interaction mediates an interaction between the spin and an external electric field that can facilitate the single spin manipulation. 

In this paper, we give a detailed theoretical investigation to the mechanism of the hyperfine interaction mediated EDSR. We emphasize the importance of FM to the driving electric field in this hyperfine mechanism. The spin-flip probability is expressed by a summation over a large number of periodical functions such that there is no Rabi oscillation for the spin-flip probability. Because the hyperfine field gives an inhomogeneous broadening to the spin splitting, the fixed-frequency driving almost cannot flip the electron spin. FM of the driving field can greatly improve the spin-flip efficiency; there is approximately a 50\% improvement in the spin-flip probability.  Also, the width of hyperfine field fluctuation sets a lower bound on the modulation amplitude. Our theory is in qualitatively good agreement with the experimental observations.

\section*{Acknowledgements}
This work is supported by National Natural Science Foundation of China Grant No.~11404020 and Postdoctoral Science Foundation of China Grant No.~2014M560039.

\appendix
\section{\label{Appendix_A}The spin-flip probability for the initial pure state $|\Psi_{00\chi_{m}}\rangle$}
\begin{figure}
\centering
\includegraphics{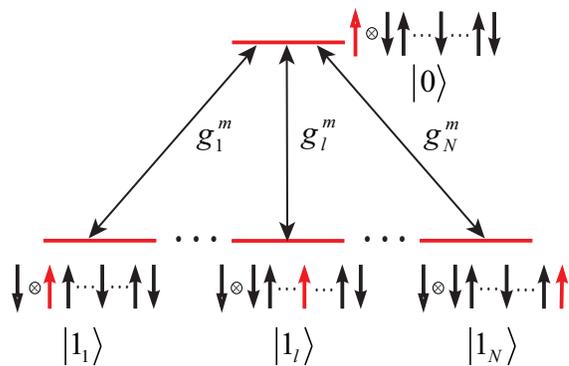}
\caption{\label{Fig_Rabiadditive}Schematically shown the simplified level diagram of the electron-nuclear system under the electric driving. There is an electric-dipole transition element $g^{m}_{l}$ between the state $|0\rangle$ and the state $|1_{l}\rangle$.}
\end{figure}

In the following, for brevity, we make some abbreviations such as $|0\rangle\equiv|\Psi_{00\chi_{m}}\rangle$ and $|1_{l}\rangle\equiv\psi_{0}(x)|1\rangle|\sigma^{m}_{1}\cdots\sigma^{m}_{l-1}0\sigma^{m}_{l+1}\cdots\sigma^{m}_{N}\rangle$. Under fixed-frequency driving $eEx\cos(\nu_{0}t)$, there is a transition element $g^{m}_{l}$ between the state $|0\rangle$ and the state $|1_{l}\rangle$ (see figure~\ref{Fig_Rabiadditive}). We have the following total Hamiltonian
\begin{equation}
H=\hbar\omega\sum^{N}_{l=1}|1_{l}\rangle\langle\,1_{l}|+\sum^{N}_{l=1}g^{m}_{l}(|0\rangle\langle\,1_{l}|+|1_{l}\rangle\langle\,0|)\cos(\nu_{0}\,t),
\end{equation}
where $\hbar\omega=-\gamma_{e}B-\sum^{N}_{l=1}(A_{l}/2)(-1)^{\sigma^{m}_{l}}$ and the transition amplitude $g^{m}_{l}$ is defined in equation~(\ref{Eq_transitionrate}). It should be noted that we have set the energy of state $|0\rangle$ to 0. The frequency of the driving electric field is in resonance with the Zeeman splitting, i.e., $\gamma_{e}B=\hbar\nu_{0}$. In the interaction picture, the Hamiltonian reads (under the rotating-wave approximation)
\begin{equation}
H_{\rm int}=\Delta_{m}\sum^{N}_{l=1}|1_{l}\rangle\langle\,1_{l}|+\sum^{N}_{l=1}\frac{g^{m}_{l}}{2}(|0\rangle\langle\,1_{l}|+|1_{l}\rangle\langle\,0|),
\end{equation}
where
\begin{equation}
\Delta_{m}=-\gamma_{e}B-\sum^{N}_{l=1}(A_{l}/2)(-1)^{\sigma^{m}_{l}}+\hbar\nu_{0}\nonumber
\end{equation}
is the detuning function. The system is initially in state $|\phi(0)\rangle=|0\rangle$, i.e., the electron spin is in the spin-up state. We want to find the spin-down probability $P_{\downarrow_{x}}(t)$ under electric driving. We let $|\phi(t)\rangle=c_{0}(t)|0\rangle+\sum^{N}_{l=1}c_{l}(t)|1_{l}\rangle$, where $c_{l}(t)$ ( $l=0,1\ldots\,N$) is the coefficient to be determined. The equations of motion for the coefficients can be derived from the Schr\"{o}dinger equation $i\hbar\partial_{t}|\phi(t)\rangle=H_{\rm int}|\phi(t)\rangle$. We find that the coefficients satisfy
\begin{eqnarray}
\dot{c}_{0}(t)&=&-i\sum^{N}_{l=1}\frac{g^{m}_{l}}{2\hbar}c_{l}(t),\nonumber\\
\dot{c}_{l}(t)&=&-i\frac{\Delta_{m}}{\hbar}\,c_{l}(t)-i\frac{g^{m}_{l}}{2\hbar}c_{0}(t).
\end{eqnarray}
The solution to this equation array is very simple, we obtain
\begin{eqnarray}
c_{0}(t)&=&e^{-i\frac{\Delta_{m}}{2\hbar}t}\left[\cos\frac{\Omega_{m}t}{2}+i\frac{\Delta_{m}}{\hbar\Omega_{m}}\sin\frac{\Omega_{m}t}{2}\right],\nonumber\\
c_{l}(t)&=&-i\frac{g^{m}_{l}}{\hbar\Omega_{m}}e^{-i\frac{\Delta_{m}}{2\hbar}t}\sin\frac{\Omega_{m}t}{2},
\end{eqnarray}
where the Rabi frequency is defined as
\begin{equation}
\Omega_{m}=\frac{1}{\hbar}\sqrt{\Delta^{2}_{m}+\sum^{N}_{l=1}(g^{m}_{l})^{2}}.
\end{equation}
Therefore, the spin down probability is given by
\begin{equation}
P_{\downarrow_{x}}(t)=\sum^{N}_{l=1}|c_{l}(t)|^{2}=\frac{\sum_{l}(g^{m}_{l})^{2}}{\hbar^{2}\Omega^{2}_{m}}\sin^{2}\frac{\Omega_{m}t}{2}.
\end{equation}
So we derive the result given in equation~(\ref{Eq_inversion_purestate}).

\end{document}